\documentstyle[11pt,newpaspiaus217,twoside,epsf]{article}
\def\edcomment#1{\iffalse\marginpar{\raggedright\sl#1\/}\else\relax\fi}
\reversemarginpar

\begin{document}
\title{Environmental Effects on the Kinematics of Virgo Cluster Galaxies}
 \author{L. Chemin$^{1,2}$, V. Cayatte$^2$, C. Balkowski$^2$, P. Amram$^3$,  M. Marcelin$^3$, O. Garrido$^3$, J. Boulesteix$^3$, 
 C. Carignan$^1$, A. Boselli$^3$, B. Vollmer$^4$, C. Adami$^3$, O. Hernandez$^{1,3}$}
 \small
\affil{$^1$ Dept. de Physique, Universit\'e de Montr\'eal, C.P. 6128, Succ. Centre-ville, Montr\'eal, Qc, Canada, H3C3J7}
\affil{$^2$ GEPI, Observatoire de Paris, 5 Pl. J. Janssen, 92195 Meudon, France}
\affil{$^3$ Observatoire Astronomique de Marseille-Provence, 2 Pl. Le Verrier, 13248 Marseille, France}  
\affil{$^4$ CDS, Observatoire de Strasbourg, 11 rue de l'Universit\'e, 67000 Strasbourg, France}  
\normalsize
\begin{abstract}
We present results from an ongoing survey dedicated to the ionized gas kinematics 
of Virgo cluster spiral galaxies using Fabry-Perot (\small FP\normalsize) interferometry. Our goal is to study 
the environmental effects on galaxy evolution in the Virgo cluster. We report here on the H$\alpha$ 
distribution map  and velocity field of NGC 4438, the prototype of an interacting galaxy near the centre of the
cluster.
\end{abstract}

\vspace*{-1.0cm}
\section{Introduction}
\vspace*{-0.3cm}
Environmental effects play an important role on the evolution of galaxies in
clusters. Different mechanisms are invoked to explain the morphology segregation observed in clusters 
like tidal interactions between galaxies and ram pressure stripping exerted by the intra-cluster medium on
disc galaxies (see Combes 2003, van Gorkom 2003 and references therein). 
 
We are completing a survey of Virgo cluster spiral galaxies by \small FP\normalsize\ observations (Chemin 2003)
in order to determine the role of tidal interactions and ram-pressure stripping 
(\small ICM-ISM\normalsize\ interactions) on their ionized gas kinematics.  
Observed data-cubes will be compared with the results of 3D simulations of \small ICM-ISM\normalsize\ 
and/or tidal interactions (eg Vollmer 2003). 

We will also investigate the influence of the environmental effects on the dark 
matter distribution by comparing mass models with the results of \small GHASP\normalsize, 
a \small FP\normalsize\ survey of local isolated galaxies (Garrido et al. 2002). 
\vspace*{-0.40cm}
\section{Observations}
\vspace*{-0.3cm}
 
We   observe   galaxies for which exist deep H$\alpha$ images (Koopmann et al. 2001, Gavazzi et al. 2002) 
and which present perturbations in their rotation curves as derived from long-slit measurements 
(Rubin et al. 1999). The survey spans all morphological types from S0a to Irregulars. 
We have already observed 25 objects which are distributed over the whole cluster, with a concentration 
around M87 in the cluster core where the environmental effects are expected to be stronger.
The observations take place at the 1.6m Observatoire du mont M\'egantic (Canada), 
1.93m Observatoire de Haute-Provence (France), 3.6m Canada-France-Hawaii and 3.6m \small ESO\normalsize\ telescopes, 
equipped with a scanning \small FP\normalsize\ interferometer coupled with a photon-counting camera of very high sensitivity 
(\small FaNTOmM\normalsize\ and \small GHASP\normalsize\ instruments, Gach et al. 2002).  

The observation of NGC 4438 (Fig.~1) has been done on April 2002 at the 3.6m \small ESO\normalsize\ telescope. The field of view 
is 3.6\arcmin\ and the spatial sampling is 0.42\arcsec. The spectral step and the total exposure time were 
16 km/s and 9360s respectively. 

\vspace*{-0.6cm}
\section{Results}
\begin{figure}
\plotone{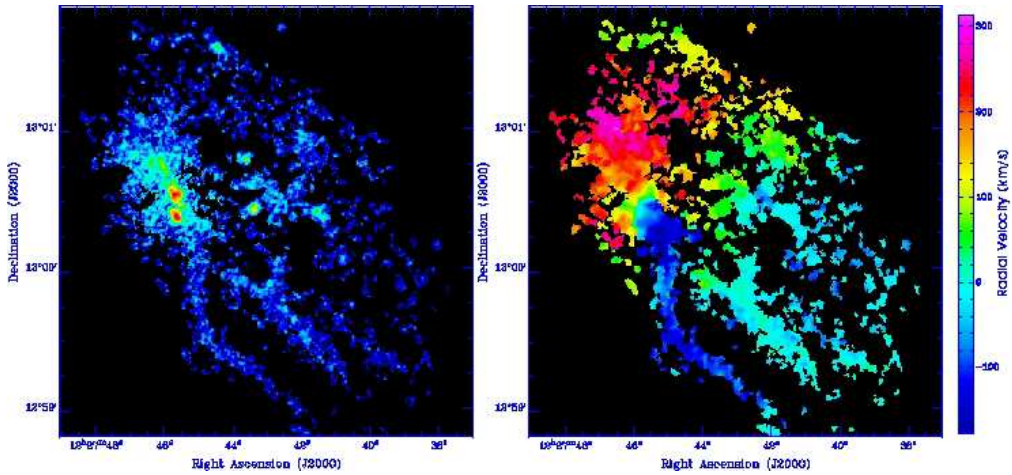}
\caption{H$\alpha$ monochromatic map (left) and velocity field (right) of NGC 4438 as obtained by \small FP\normalsize\ observations.}
\end{figure}

\vspace*{-0.25cm}
Figure~1 (left) presents the H$\alpha$ emission map of NGC 4438, a highly inclined disc located near the cluster core. 
It shows bright H\small II\normalsize\ regions which are associated to the galactic disc and to the nuclear outflow recently evidenced by Kenney \& Yale (2002).   
Diffuse filaments along which are observed a few brighter clumps are seen on the western side of the disc. 
The right panel of the figure presents the complex velocity field of NGC 4438 (Chemin et al., in prep.).   
Both fields indicate strong signs of perturbations, probably induced by an \small ICM-ISM\normalsize\  interaction 
and/or a tidal interaction with a companion.

\end{document}